\documentclass[prc,twocolumn,showpacs,preprintnumbers,amsmath,amssymb,superscriptaddress,floatfix,nofootinbib]{revtex4-1}
\usepackage{subfigure,dcolumn}
\usepackage{epstopdf}
\usepackage{mhchem}
\usepackage{upgreek}

\usepackage{graphicx}
\usepackage{color}
\usepackage{CJK}
\usepackage{ulem}
\newcommand{\beq}{\begin{equation}}
\newcommand{\eeq}{\end{equation}}
\newcommand{\beqn}{\begin{eqnarray}}
\newcommand{\eeqn}{\end{eqnarray}}

\usepackage{amsmath}
\allowdisplaybreaks[4]

\renewcommand\sout{\bgroup \color{red} \ULdepth=-.5ex \ULset}

\renewcommand{\rm}[1]{\textrm{#1}}
\renewcommand{\d}{\mathrm{d}}

\def\esym{$E_{\rm{sym}}(\rho)$~}

\def\es0{$E_{\rm{sym}}(\rho_0)$}

\renewcommand{\rm}[1]{\textrm{#1}}
\renewcommand{\d}{\mathrm{d}}

\begin{document}
\title{How tightly is the nuclear symmetry energy constrained by a unitary Fermi gas?}
\author{Nai-Bo Zhang}
\affiliation{Department of Physics and Astronomy, Texas A\&M University-Commerce, Commerce, TX 75429, USA}
\affiliation{Shandong Provincial Key Laboratory of Optical Astronomy and Solar-Terrestrial
Environment, Institute of Space Sciences, Shandong University, Weihai 264209, China}
\author{Bao-Jun Cai}
\affiliation{Department of Physics and Astronomy, Texas A\&M
University-Commerce, Commerce, TX 75429, USA}
\affiliation{Department of Physics, Shanghai University, Shanghai
200444, China} \affiliation{Department of Physics and Astronomy and
Shanghai Key Laboratory for Particle Physics and Cosmology, Shanghai
Jiao Tong University, Shanghai 200240, China}
\author{Bao-An Li\footnote{Corresponding author: Bao-An.Li@Tamuc.edu}}
\author{William G. Newton}
\affiliation{Department of Physics and Astronomy, Texas A\&M University-Commerce, Commerce, TX 75429, USA}
\author{Jun Xu}
\affiliation{Shanghai Institute of Applied Physics, Chinese Academy of Sciences, Shanghai 201800, China}
\date{\today}

\begin{abstract}
We examine critically how tightly the density dependence of nuclear symmetry energy \esym is constrained by
the universal equation of state (EOS) of the unitary Fermi gas
$E_{\rm{UG}}(\rho)$ considering currently known uncertainties of higher order parameters describing the density dependence of the Equation of State of isospin-asymmetric nuclear matter.
We found that $E_{\rm{UG}}(\rho)$ does provide a useful
lower boundary for the \esym. However, it does not tightly constrain the
correlation between the magnitude $E_{\rm{sym}}(\rho_0)$ and slope
$L$ unless the curvature $K_{\rm{sym}}$ of the symmetry energy at
saturation density $\rho_0$ is more precisely known. The large uncertainty in
the skewness parameters affects the  $E_{\rm{sym}}(\rho_0)$ versus
$L$ correlation by the same almost as significantly as the uncertainty in $K_{\rm{sym}}$.
\end{abstract}

\maketitle

\noindent{\it Introduction:} To understand the nature of
neutron-rich nucleonic matter has been a major scientific goal in
both nuclear physics and astrophysics. The density dependence of
nuclear symmetry energy \esym has been a major uncertain part of
the equation of state (EOS) of neutron-rich matter especially at
high densities, see, e.g., collections in\,\cite{Tesym}. Reliable knowledge about the \esym has
significant ramifications in answering many interesting questions
regarding the structure of rare isotopes and neutron stars, dynamics
of heavy-ion collisions and supernova explosions as well as the
frequency and strain amplitude of gravitational waves from deformed
pulsars and/or cosmic collisions involving neutron stars. During the
last two decades, significant efforts have been devoted to exploring
the \esym using both terrestrial laboratory
experiments\,\cite{ireview98,ibook01,ditoro,LCK08,Lynch09,Dan09,Trau12,Tsang12,Chuck14,Baldo16,Li2017}
and astrophysical
observations\,\cite{Steiner05,Lat13,Newton14,Lida14,Pearson,Far14,Fis14,Blas16}.
Extensive surveys of the extracted constraints on the \esym around
the saturation density $\rho_0$ indicate that the central values of
the $E_{\rm{sym}}(\rho_0)$ and its slope $ L=\left[3 \rho (\partial
E_{\rm{sym}}/\partial \rho\right)]_{\rho_0} $ scatter around $31.6$
MeV and $58.9$ MeV, respectively\,\cite{Lat13,Li-han,Oer16}. At
densities away from $\rho_0$, however, the \esym remains rather
unconstrained especially at supra-saturation
densities\,\cite{Li2017}.

Interestingly, recent progresses in another seemingly different
field may provide additional information about the density
dependence of nuclear symmetry energy. Indeed, theoretical and
experimental studies of cold atoms have made impressive progress in
recent years, see, e.g., refs.\,\cite{Zwi15,Ku12,Zu13,Endres13} for recent reviews,
providing reliable information about the universal EOS ($E_{\rm{UG}}$)
of unitary gas (UG) interacting via pairwise $s$-waves with
infinite scattering length but zero effective range. The universal
$E_{\rm{UG}}$ constrains stringently the EOS of pure neutron
matter (PNM) at sub-saturation densities, thus provides possibly
additional constraints on the nuclear symmetry energy. In fact, it
was recently conjectured that the $E_{\rm{UG}}$ provides the lower
boundary of the EOS of PNM ($E_{\rm{PNM}}$) \cite{Kolomeitsev16}. Moreover,
using a set of known parameters of symmetric nuclear matter (SNM), and taking zero as an 
upper bound on the curvature $K_{\rm{sym}}=\left[9\rho^2(\partial^2E_{\rm{sym}}(\rho)/\partial\rho^2)\right]_{\rho_0}$
of \esym at $\rho_0$, the authors of ref.\,\cite{Kolomeitsev16} obtained a region of $E_{\rm{sym}}(\rho_0)$ - $L$ space that is inconsistent with the unitary gas constraints, excluding many \esym functionals currently actively used in both nuclear physics and astrophysics.

Our original purposes were to examine several issues not clearly addressed in version-1 of ref. \cite{Kolomeitsev16}. We notice that some of these issues are now discussed in more detail in its revised version. 
Nevertheless, it is still useful to provide our results and opinions on some of these issues. 
The derivation of the excluded region in $E_{\rm{sym}}(\rho_0)$ - $L$ space by ref. \cite{Kolomeitsev16} relies on several assumptions \cite{NST}: 
{\it the most importance of which is the underlying conjecture that $E_{\rm{PNM}}(\rho)\geq E_{\rm{UG}}(\rho)$ where $E_{\rm{UG}}(\rho)=\xi E_F(\rho)$ with $\xi$ being the Bertsch parameter \cite{Zwi15,Ku12,Zu13,Endres13}
and $E_F\propto (\rho/\rho_0)^{2/3}$ is the energy of a non-interacting non-relativistic degenerate Fermi gas of neutrons. This conjecture is not quite the same as stating that PNM has a greater energy than the UG at all densities, since the UG is experimentally accessible only at low densities. This conjecture merely employs an algebraic expression motivated by the result that apparently the low-density neutron gas has a higher energy
than the UG, and by the inference that the repulsive nature of the three nucleon (NNN) interaction coupled with the finite values of the range and scattering length of the two-body (NN) s-wave interaction will ensure that the energy of PNM remains above this algebraic expression at higher densities.  A further assumption is that possibly attractive higher order NN interactions (p-wave, d-wave, etc.) are not important relative to the repulsive character of three- and higher-body interactions. Additionally, neutrons are assumed to remain non-relativisitc in the density range considered.}
Given the important ramifications of the findings in ref. \cite{Kolomeitsev16}, we are motivated to critically examine them adopting the same assumptions. We include the third-order terms in density characterized by the skewness coefficients $J_0=27\rho_0^3\partial^3E_0(\rho)/\partial\rho^3|_{\rho=\rho_0} $ and $J_{\rm{sym}}=27\rho_0^3\partial^3E_{\rm{sym}}(\rho)/\partial\rho^3|_{\rho=\rho_0}$ in expanding the $E_{\rm{0}}(\rho)$ and $E_{\rm{sym}}(\rho)$, respectively. Particularly, we carefully examine the uncertainties of the curvature $K_{\rm{sym}}$ of the symmetry energy and the skewness coefficients 
$J_0$ and $J_{\rm{sym}}$, taking into account energy density functionals that are consistent with the PNM EOS derived from microscopic calculations, and examine the effects of those uncertainties on the region of $E_{\rm{sym}}(\rho_0)$-$L$ space excluded by the unitary gas constraints.

While Skyrme models consistent with microscopic PNM calculations tend to give $K_{\rm{sym}}$ in the range -100 to -200 MeV \cite{Fattoyev12,Brown13}, Relativistic Mean Field (RMF) models consistent with microscopic PNM calculations can give positive values of $K_{\rm{sym}}$ \cite{Fattoyev12,Dutra14}, reflecting a difference in the form of these two classes of energy density functionals. Indeed, some reputable non-relativistic and relativistic energy density functionals in the literature, see, e.g., reviews in ref.\ \cite{Lwchen09,Dutra12,Dutra14,Colo14}, predict positive $K_{\rm{sym}}$ values and meet all existing constraints including the EOS of PNM within their known uncertain ranges. For example, the TM2 RMF interaction has $K_{\rm{sym}}$ = 50 MeV and passes the PNM test of ref.\cite{Dutra14}. To our best knowledge, while the majority of existing models predict negative values for $K_{\rm{sym}}$, there is no fundamental physics principle excluding positive $K_{\rm{sym}}$ values. The current situation clearly calls for more studies on the $K_{\rm{sym}}$ especially its experimental constraints. Hopefully, ongoing experiments at several laboratories \cite{Umesh} to extract the isospin dependence of nuclear incompressibility and subsequently the $K_{\rm{sym}}$ from giant resonances of neutron-rich nuclei will help settle the issue in the near future. 
Interestingly, very recently by analyzing comprehensively the relative elliptical flows of neutrons and protons measured by the ASY-EOS and the FOPI-LAND collaborations at GSI using a Quantum Molecular Dynamics (QMD) 
model \cite{Cozma}, the extracted values for the slope and curvature parameters are $L=59\pm 24$ MeV and $K_{\rm sym}=88\pm 372$ MeV at the $1\sigma$ confidence level. After considering uncertainties of other model parameters including the incompressibility of SNM, neutron-proton effective mass splitting (related to the momentum dependence of the symmetry potential), Pauli blocking and in-medium nucleon-nucleon cross sections, it was concluded that $L = 59 \pm 24 (\rm{exp}) \pm 16(\rm{th}) \pm10 (\rm{sys})$ MeV and 
$K_{\rm{sym}} = 0 \pm 370(\rm{exp}) \pm 220(\rm{th}) \pm 150(\rm{sys})$ MeV. To our best knowledge, the latter represents the latest and most accurate constraint on $K_{\rm{sym}}$. 
\\


\begin{figure}[t]
  \centering
  \includegraphics[width=8cm]{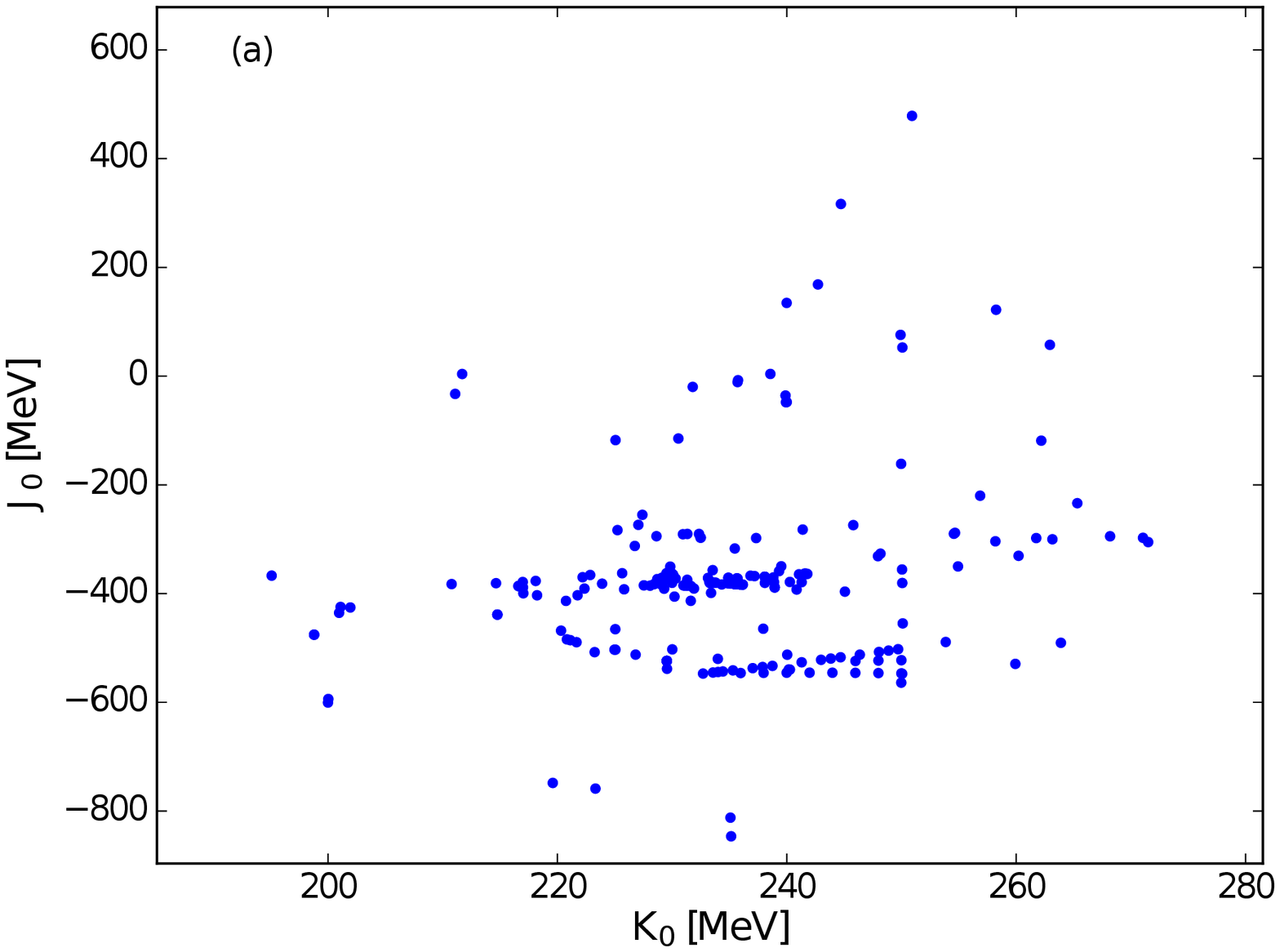}\\
   \includegraphics[width=8cm]{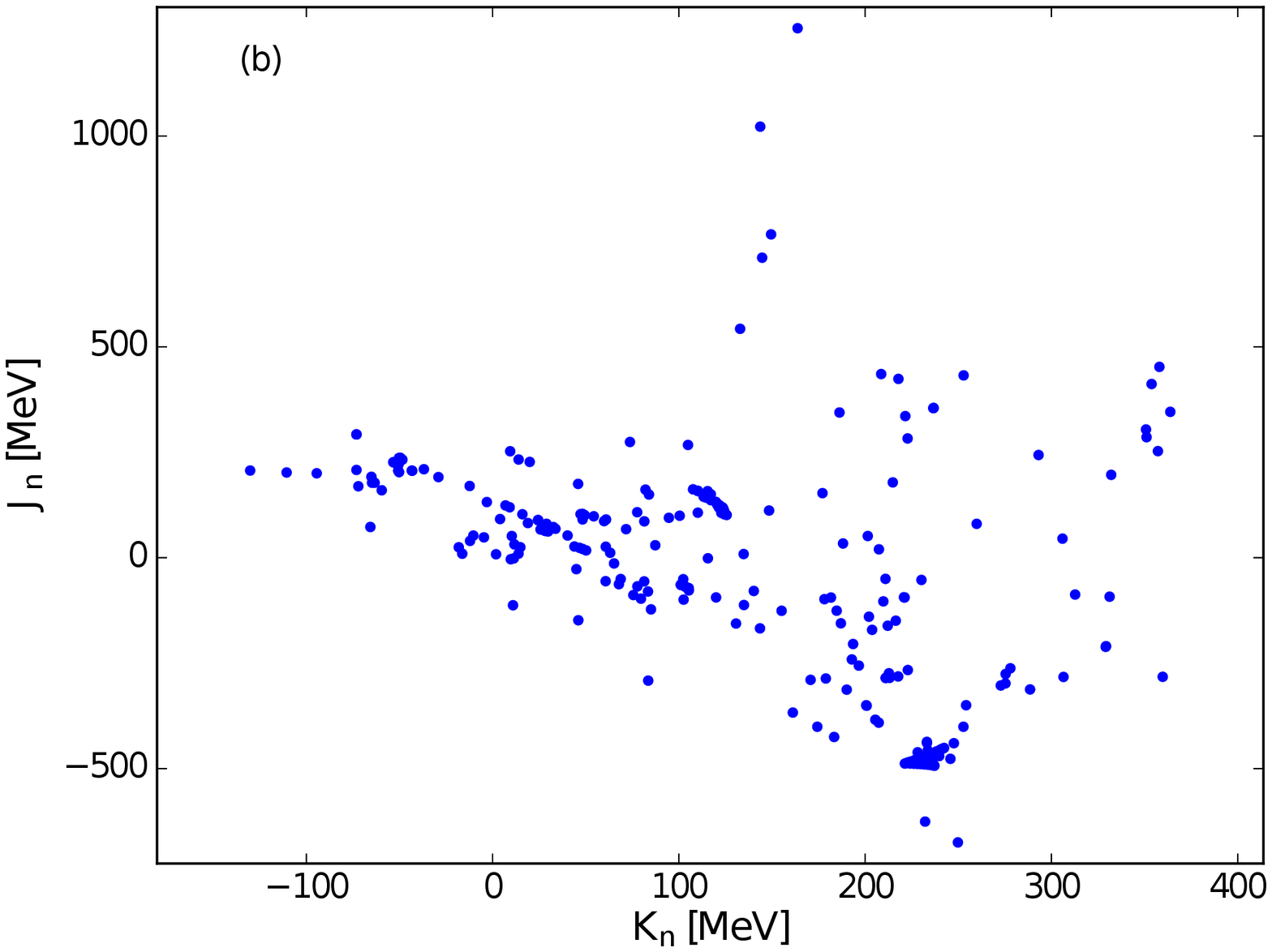}\\
  \caption{(Color online) The skewness parameter $J_0$ versus the incompressibility $K_0$ for symmetric matter (a) and the total curvature parameter $K_{\rm n} = K_0 + K_{\rm sym}$ versus the total skewness parameter $J_{\rm n} = J_0 + J_{\rm sym}$ (b) for all 173 Skyrme and 101 RMF models examined by Dutra \textit{et al} \cite{Dutra12,Dutra14} which pass their pure neutron matter constraints and satisfy 190$<K_0<$270 MeV.}\label{KJ}
\end{figure}



\noindent{\it The lower boundary of nuclear symmetry energy
constrained by the universal EOS of unitary Fermi gas:} Within the
parabolic approximation for the EOS of isospin asymmetric nuclear
matter (ANM) in terms of the energy per nucleon $E$, i.e., $ E(\rho
,\delta )=E_0(\rho)+E_{\rm{sym}}(\rho )\delta ^{2}
+\mathcal{O}(\delta^4), $
 the symmetry energy $E_{\rm{sym}}(\rho )= 2^{-1}[\partial ^{2}E(\rho,\delta )/\partial \delta ^{2}]_{\delta =0}$ can be approximated by
 \begin{equation}\label{Esymps}
  E_{\rm{sym}}(\mu)\approx E_{\rm{PNM}}(\mu)-E_{\rm{0}}(\mu),
\end{equation}
where $\mu=\rho/\rho_0$ is the reduced density and
$\delta=(\rho_{\rm{n}}-\rho_{\rm{p}})/\rho$ is the isospin asymmetry
of ANM. Using the conjecture $E_{\rm{PNM}}(\rho)\geq
E_{\rm{UG}}(\rho)$\,\cite{Kolomeitsev16} and the EOS of unitary gas
\beq
E_{\rm{UG}}(\mu)=\frac{3\hbar^2k_{\rm{F}}^2}{10m_{\rm{n}}}\xi\equiv
E^0_{\rm{UG}}\mu^{2/3} \eeq where $k_{\rm{F}}$ is the neutron Fermi
momentum, the lower boundary of
symmetry energy can be obtained from
\begin{equation}\label{Esymlb}
E_{\rm{sym}}(\mu)\geq E_{\rm{UG}}(\mu)-E_{\rm{0}}(\mu)=E^0_{\rm{UG}}\mu^{2/3}-E_{\rm{0}}(\mu).
\end{equation}
It is necessary to caution that the above lower boundary of $E_{\rm{sym}}(\mu)$ is estimated based on the conjecture $E_{\rm{PNM}}(\rho)\geq E_{\rm{UG}}(\rho)$ and the assumption that $\xi$ is 
a constant in the density range we study.  As emphasized in ref. \cite{Kolomeitsev16}, the conjecture is empirical in nature. While there are strong supports for the conjecture 
by comparing the EOSs of PNM calculated from various microscopic many-body theories with the $E_{\rm{UG}}(\mu)$ using a constant $\xi\approx 0.37$ (see Fig.1 of ref. \cite{Kolomeitsev16} and Fig.2 of ref. \cite{CaiLi15}) up to 
about $\rho_0$, the rigorous condition for unitarity is expected to be reached in PNM only at very low densities. Although one can not prove the validity of the conjecture at high densities, strong 
physical arguments were made to justify and use it up to about $1.5\rho_0$ in ref. \cite{Kolomeitsev16}.
Thus, the results of our study should be understood with the caveat that they are obtained under the above reasonable but not rigorously proven conjecture and assumptions. Nevertheless, they are 
useful for comparing with the results of ref. \cite{Kolomeitsev16} obtained using the same assumptions. 

The EOS of SNM around $\rho_0$ can be expanded to the third order in density as
\begin{equation}\label{3density}
E_{\rm{0}}(\mu)=E_0(\rho_0)+\frac{K_0}{18}(\mu-1)^2+\frac{J_0}{162}(\mu-1)^3+\mathcal{O}[(\mu-1)^4]
\end{equation}
in terms of the incompressibility $K_0$ and skewness $J_0$. At the
saturation point of SNM, we adopt $E_0(\rho_0) = -15.9$\,MeV and
$\rho_0 = 0.164$\,fm$^{-3}$\,\cite{Brown13}. The lower boundary of
\esym thus depends on the values of $K_0$, $J_0$ and $\xi$. 

The incompressibility $K_0$ of SNM has been extensively
investigated\,\cite{Colo14,Stone14}, and the most widely used values are
$K_0 = 240\pm20$\,MeV\,\cite{Shlomo06,Piekarewicz10} or
$230\pm40$\,MeV\,\cite{Khan12}. However, the skewness coefficient
$J_0$ is still poorly
known\,\cite{Meixner13,Chen11,Farine97,Klahn06,Mas16,Cai14}. In Fig. \ref{KJ}(a), we show $K_0$ and $J_0$ from 
274 parameterizations of the Skyrme and RMF models that pass the PNM tests of Dutra \textit{et al} \cite{Dutra12,Dutra14} and satisfy $K_0 = 230\pm40$\,MeV. The spread in values for $J_0$ is very large, covering the range $\approx -800<J_0<$400 MeV.

\begin{figure}[t]
 \centering
  \includegraphics[width=8cm]{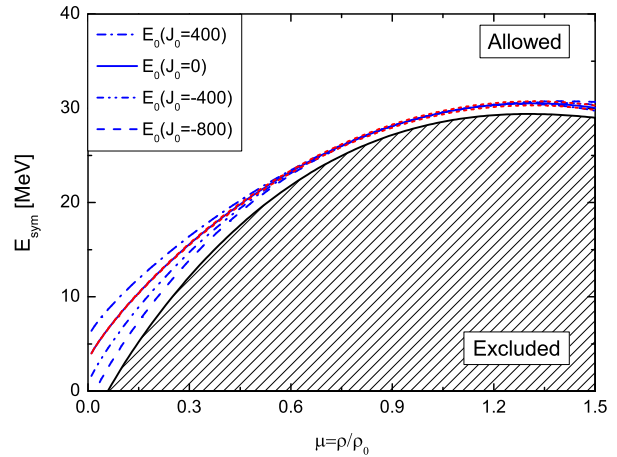}\\
  \caption{(Color online) The lower boundary of symmetry energy as a function of density for different skewness coefficients $J_0 = 400$, $0$, $-400$, and $-800$\,MeV.
  The red dashed region represents the variation of $E_{\rm{sym}}(\mu)$ with $K_0 = 230$\,MeV and $J_0 = 0$\,MeV by adopting $\xi = 0.37\pm0.005$.
  The shadowed region shows the excluded region after considering the uncertainties of $\xi$, $K$ and $J_0$.}\label{Esymrho}
\end{figure}

As reviewed recently in refs.\,\cite{Endres13,Kolomeitsev16},
currently the best estimate for the Bertsch parameter $\xi$ from
lattice Monte Carlo studies is $\xi = 0.372(5)$ consistent with the
most accurate experimental value of $\xi = 0.376(4)$. While its values from various other models and experiments have scattered
between 0.279 and 0.449(9) within the last decade, it appears that it now has converged to $\xi = 0.37\pm0.005$ which we adopt in this work.
Shown in Fig. \ref{Esymrho} with the red dashed lines are the variation of \esym
with $\xi = 0.37\pm0.005$, $J_0 = 0$\, and $K=230$\,MeV. 
Effects of varying the $\xi$ value are very small within the range considered.

Secondly, effects of the skewness parameter are shown by varying the value of
$J_0$ between $-800$ MeV and $400$ MeV, the range covered by the models plotted in Fig. \ref{KJ}(a). Although the range is very large, it translates to a range of uncertainty for \esym that is equivalent to the range of uncertainty in $K_0$ (80 MeV). This
is easy to understand as the expansion of SNM's EOS
converges quickly around the normal density by design (the $J_0$
contribution of $J_0/162$ is a factor of 9 less than the
$K_0/18$ term). 

Considering the uncertainties of all relevant parameters involved, the most
conservative lower boundary of \esym shown as the shadowed region in
Fig. \ref{Esymrho} is obtained by using $ \xi=0.37$, 
$\rho_0=0.157$ fm$^{-3}$, $E_0(\rho_0)=-15.5$ MeV, and $K_0$=270 MeV; for $\mu\leq 1$, $J_0=-800$\,MeV  and for $\mu>1$,  $J_0=400$ MeV. Overall, our observations and results are consistent with the 
findings in ref.\,\cite{Kolomeitsev16}.


\noindent{\it Constraining the $E_{\rm{sym}}(\rho_0)$ versus
$L$ boundary:} The symmetry energy $E_{\rm{sym}}(\mu)$ can
be expanded around $\rho_0$ to third order in density as
\begin{align}\label{Su}
  E_{\rm{sym}}(\mu)=&E_{\rm{sym}}(\rho_0)+\frac{L}{3}(\mu-1)+\frac{K_{\rm{sym}}}{18}(\mu-1)^2\nonumber\\
  &+\frac{J_{\rm{sym}}}{162}[(\mu-1)^3]+\mathcal{O}[(\mu-1)^4]
\end{align}
 in terms of its magnitude $E_{\rm{sym}}(\rho_0)$, slope $L$, curvature $K_{\rm{sym}}$ and skewness $J_{\rm{sym}}$ at $\rho_0$.
Inserting the above equation into Eq. (\ref{Esymlb}), the lower
boundary of $E_{\rm{sym}}(\rho_0)$ can be expressed as
\begin{align}\label{relb3}
  E_{\rm{sym}}(\rho_0)\geq& E_{\rm{UG}}^0\mu^{2/3}-E_0(\rho_0)-\frac{L}{3}(\mu-1)-\frac{K_{n}}{18}(\mu-1)^2\nonumber\\
  &-\frac{J_n}{162}(\mu-1)^3
\end{align}
where $K_n=K_{\rm{sym}}+K_0$ and $J_n=J_{\rm{sym}}+J_0$. Taking the derivative of the above
equation with respect to density on both sides, one can readily get
an expression for the lower boundary of $L$
\begin{equation}\label{relbd3}
L=\frac{2E_{\rm{UG}}^0}{\mu^{1/3}}-\frac{K_{n}}{3}(\mu-1)-\frac{J_n}{18}(\mu-1)^2.
\end{equation}
Then, putting the above expression back to Eq. (\ref{relb3}) the
latter can be rewritten as
\begin{align}\label{relbj}
  E_{\rm{sym}}(\rho_0)\geq& \frac{E_{\rm{UG}}^0}{3\mu^{1/3}}(\mu+2)+\frac{K_{n}}{18}(\mu-1)^2\nonumber\\
  &+\frac{J_n}{81}(\mu-1)^3-E_0(\rho_0).
\end{align}
These two equations reveal the correlation between the $E_{\rm{sym}}(\rho_0)$ and $L$ along their
lower boundaries through the arbitrary density $\mu$. Setting $J_n$=0, the Eqs. (\ref{relbd3}) and (\ref{relbj}) reduce
exactly to the parametric equations of $E_{\rm{sym}}(\rho_0)$ and $L$ derived slightly differently in version-1 of ref.\,\cite{Kolomeitsev16}. We note that the quantities that determine the boundary of allowed values of $E_{\rm sym}(\rho_0)$ and $L$ are  the total curvature parameter $K_n $ and total skewness parameter $J_n$ also emphasized in version-2 of ref.\,\cite{Kolomeitsev16}.

While having noted that $K_{\rm{sym}}$ is experimentally and
theoretically poorly known, the $E_{\rm{sym}}(\rho_0)$ versus $L$
correlation along their boundaries was obtained in ref. \cite{Kolomeitsev16} by setting
$K_{\rm{sym}}=0$ based on the prediction of a chiral effective field theory. 
It was found that the resulting correlation excludes many of the currently actively
used models for \esym. We reexamine this correlation by varying the $\xi$, $J_n$ and $K_{\rm{sym}}$
within their known uncertain ranges. Again, the value of $\xi$ is now well settled around $0.37\pm 0.005$. Taking
$K_{\rm{sym}}=0$, $J_n=0$ and $K_0=230$ MeV, the two red dashed lines
obtained with $\xi = 0.37\pm 0.005$ in both (a) and (b) of Fig. 3 show
the resulting lower boundaries of the $E_{\rm{sym}}(\rho_0)$ versus
$L$ correlation. 

The skewness coefficients $J_0$ and $J_{\rm{sym}}$ in $J_n$ are both poorly known. We show in Fig \ref{KJ}(b) values of $J_n$ against $K_n$ for the 275 Skyrme and RMF models. $J_n$ varies approximately in the range $-500$ MeV $\leq J_n \leq 1000$MeV. To our best knowledge, there is no experimental constraint available on this quantity. $K_n$ varies approximately in the range $-150$ MeV $\leq K_n \leq 370$ MeV. Most of this comes from the big uncertainties in determining the value of $K_{\rm{sym}}$, which are discussed in detail in ref.\,\cite{Li2017}. This is partially because the
$K_{\rm{sym}}$ depends on not only $L$ but also its derivative $(\d L/\d \rho)_{\rho_0}$ by definition. Microscopically, it depends on not only the nucleon isoscalar effective mass $m^*_0$ and neutron-proton effective mass slitting $m_n^*-m_p^*$ but also their momentum and density dependences that are all essentially completely unknown\,\cite{Li2017}. 
The latest calculations within many Skyrme Hartree-Fock and/or relativistic mean-field models indicate that
$-400 \leq K_{\rm{sym}}\leq 100$\,MeV \cite{Lwchen09,Dutra12,Dutra14,Colo14}.

\begin{figure}[t]
\centering
  \includegraphics[width=8cm]{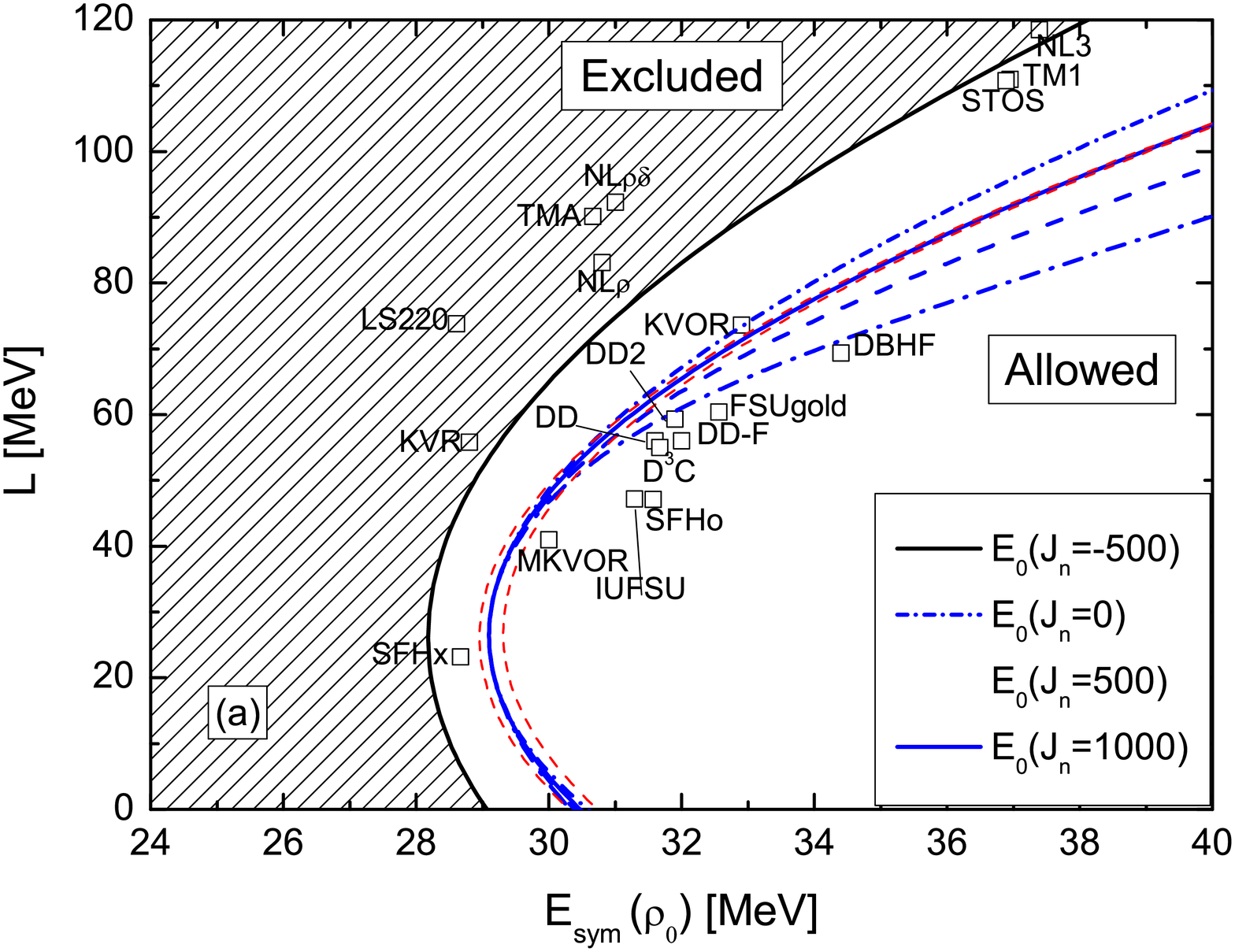}\\
  \includegraphics[width=8cm]{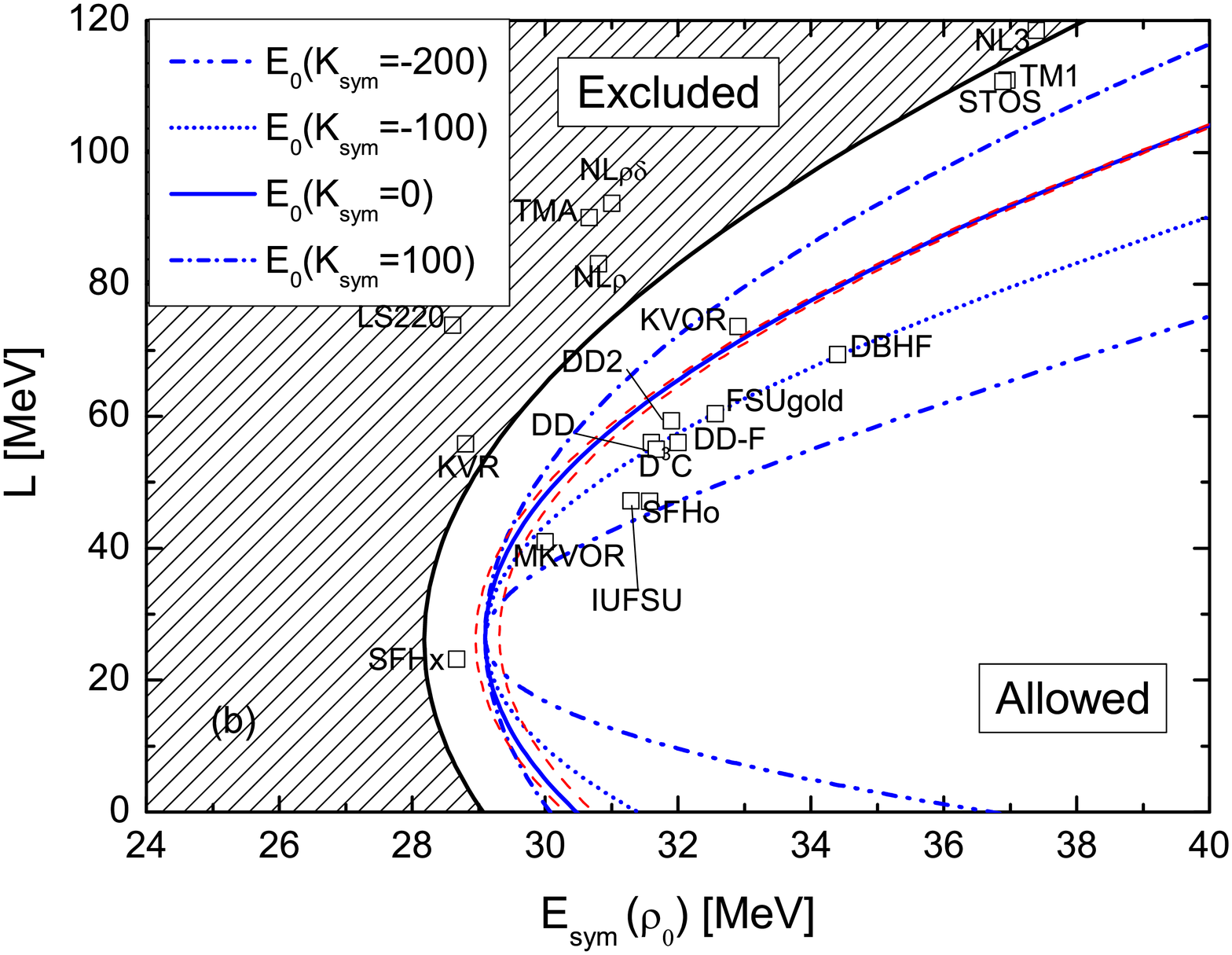}\\
  \caption{(Color online) The lower boundaries of symmetry energy parameters based on different parameter values. The effects of the uncertainty in $J_n$ (a) and $K_{\rm{sym}}$ (b) are demonstrated by the blue lines, holding $K_n=230$ MeV in (a) and $J_{\rm n}$=0 MeV in (b). The red dashed region represents the deviation of $E_{\rm{sym}}$ with $J_0 = 0$ MeV by adopting $\xi = 0.37\pm0.005$. The shadowed region shows the excluded region after considering all the uncertainties of $\rho_0$, $E_{0}(\rho_0)$, $\xi$, $K_0$, $J_n$, and $K_{\rm{sym}}(\rho_0)$.}\label{S0lho}
\end{figure}

The results using $K_{\rm{sym}}=0$, $K_0=230$ MeV (so $K_n =$ 230 MeV) and $\xi=0.37$ for
different values for $J_n$ are shown in Fig. \ref{S0lho} (a). It is
seen that $L$ becomes larger as $J_0$ \textit{decreases}, and that the upper boundary of the allowed region will correspond to the lower limit of $J_n$. At $E_{\rm{sym}}(\rho_0) = 40$\,MeV, for
example, the increase is only about $20\%$. 

By setting $K_{\rm{sym}}= -200, -100$, $0$, and $100$\,MeV with $K_0=230$ MeV (corresponding to $K_n=30, 130, 230$ and $330$ MeV) we can illustrate effects of the $K_{\rm{sym}}$ on the lower boundary of
$E_{\rm{sym}}(\rho_0)$ versus $L$ correlation in Fig. \ref{S0lho}
(b). It is seen that the $K_{\rm{sym}}$ affects the results
significantly; as $K_{\rm sym}$ (and hence $K_n$) \textit{increases}, $L$ increases, and the upper boundary of the allowed region will correspond to the largest value of $K_{\rm sym}$ (and hence $K_n$). The overall uncertainty in $K_n$ leads to a variation of the upper boundary of the allowed region that is about twice that caused by the uncertainty in $J_n$. We note that many of the models allowed by
$K_{\rm{sym}}= 0$\,MeV would be excluded by using $K_{\rm{sym}}=-200$\,MeV.


Adopting the following values for the five parameters after taking into account the full range presented by the 275 Skyrme and RMF models, i.e.,
\begin{eqnarray}\label{lbparameter}
  &&\rho_0=0.157\,{\rm{fm}}^{-3}, ~E_{0}(\rho_0)=-15.5\,{\rm{MeV}}, ~\xi=0.37, \nonumber\\
  &&K_n=370\,{\rm{MeV}},~J_n=-500\,{\rm{MeV}}\nonumber
\end{eqnarray}
a lower boundary excluding only the shadowed region in Fig.
\ref{S0lho} is obtained. It is seen that only the TMA and
NL$\rho\delta$, NL3 and LS220 may be excluded, while the STOS, TM1, NL$\rho$,
LS220, and KVR, which have been surely excluded previously in
ref.\,\cite{Kolomeitsev16} may be allowed. This is mainly a result in extending the upper bound in the uncertainty region of $K_{\rm sym}$, since the additional uncertainty in $J_{n}$ moves the excluded region to the right, (because $J_n \gtrsim 0$).  As pointed out in the final analyses of ref. \cite{Kolomeitsev16}, we agree that strong empirical correlations among $J_{\rm {sym}}, K_{\rm{sym}}$ and $L$ exist. These correlations can be used to refine the 
$E_{\rm{sym}}(\rho_0)$-L constraint shown in Fig.\ \ref{S0lho}.

It is well known that the detailed density dependence of nuclear symmetry energy \esym 
contains many interesting and some unknown physics. It is probably not surprising that the estimation of the correlation between its 
zeroth-order and first-order density expansion coefficients $E_{\rm{sym}}(\rho_0)$ and $L$ depends on what we assume about the immediate next high-order term characterized by the curvature $K_{\rm{sym}}$ at $\rho_0$. 
\\ 

\noindent{\it Concluding remarks:} The universal EOS $E_{\rm{UG}}$ of the unitary Fermi gas was conjectured in ref. \cite{Kolomeitsev16} to provide the lower boundary of the EOS $E_{\rm{PNM}}$ of PNM, 
and thus a constraint on nuclear symmetry energy. Although unproven, the conjecture has strong empirical supports, and its implications are important enough that they should be examined rigorously. 
We found that the $E_{\rm{UG}}$ does provide a useful lower boundary of nuclear symmetry energy. Moreover, this boundary is essentially not affected by the
known uncertainty in the skewness coefficient $J_0$ of SNM. However,
it does not tightly constrain the correlation between the magnitude
$E_{0}(\rho_0)$ and slope $L$ unless the curvature $K_{\rm{sym}}$ of
the \esym at $\rho_0$ and the skewness parameters $J_0$ and $J_{\rm sym}$ are 
better known. Of these, $K_{\rm{sym}}$ is the more important quantity, its uncertainty affecting the lower boundary of \esym by up
to twice as much as the uncertainty in $J_n$. Most of the previously
excluded \esym functionals by the universal EOS of unitary Fermi gas
assuming $K_{\rm{sym}}=0$ may not be excluded considering the
currently known big uncertainties of the $K_{\rm{sym}}(\rho_0)$.

For many purposes in both nuclear physics and astrophysics, it is
necessary to map out precisely both the \esym and the EOS $E_{0}(\rho)$ of SNM in
a broad density range. Near the saturation density $\rho_0$, this
requires accurate knowledge of the $K_{\rm{sym}}$ and $J_0$ besides
the $E_{0}(\rho_0)$, $L$ and $K_0$. To this end, it is interesting to
mention briefly quantities that are sensitive to the higher-order
EOS parameters and current efforts to determine them. For example,
the skewness coefficient $J_0$ characterizes the high-density
behavior of $E_{0}(\rho)$, and it has been found to affect significantly the
maximum mass of neutron stars\,\cite{Cai14}. Moreover, at the
crust-core transition point where the incompressibility of neutron
star matter at $\beta$-equilibrium vanishes, the value of $J_0$
influences significantly the exact location of the transition point
\cite{NBZ}. Thus, astrophysical observations of neutron stars can
potentially constrain the $J_0$ {\it albeit} probably not before other EOS
parameters are well determined. On the other hand, in terrestrial
laboratory experiments, there have been continued efforts to
determine the $K_{\rm{sym}}$\,\cite{Colo14}. One
outstanding example is the measurement of the isospin dependence of
nuclear incompressibility $K(\delta)\approx
K_0+K_{\tau}\delta^2+\mathcal{O}(\delta^4)$ where
$K_{\tau}=K_{\rm{sym}}-6L-J_0L/K_0$ using giant
resonances of neutron-rich nuclei\,\cite{TLi,Jorge-TLi}.
While the current estimate of $K_{\tau}\approx -550\pm 100$\,MeV\,\cite{Colo14} from
analyzing many different kinds of terrestrial experiments is still
too rough to constrain tightly the individual values of $J_0$ and
$K_{\rm{sym}} $, new experiments with more neutron-rich beams have
the promise of improving significantly the accuracy of the measured
$K_{\tau}$\,\cite{Umesh}. Thus, we are hopeful that not only the
zeroth and first-order parameters $K_0$, $E_{\rm{sym}}(\rho_0)$ and
$L$ but also high-order coefficients $J_0$ and $K_{\rm{sym}}$ can be
pinned down in the near future by combining new analyses of upcoming
astrophysical observations and terrestrial experiments.\\

\noindent{\it Acknowledgements:} We would like to thank Umesh Garg for helpful communications. NBZ is supported in part by the China Scholarship Council. BAL acknowledges the U.S. Department of Energy, Office of Science, under Award Number DE-SC0013702, the CUSTIPEN (China-U.S. Theory Institute for Physics with Exotic Nuclei) under the US Department of Energy Grant No. DE-SC0009971, the National Natural Science Foundation of China under Grant No. 11320101004 and the Texas Advanced Computing Center. JX is supported in part by the Major State Basic Research Development Program (973 Program) of China under Contract Nos. 2015CB856904 and 2014CB845401, the National Natural Science Foundation of China under Grant Nos. 11475243 and 11421505, the ``100-talent plan'' of Shanghai Institute of Applied Physics under Grant Nos. Y290061011 and Y526011011 from the Chinese Academy of Sciences, the Shanghai Key Laboratory of Particle Physics and Cosmology under Grant No. 15DZ2272100, and the Shanghai Pujiang Program under Grant No. 13PJ1410600.

\end{document}